\documentclass[prd]{revtex4}
\usepackage{graphicx}
\begin{document}

\title{New Higgs signals induced by mirror fermion mixing effects}

\author{U. Cotti$^1$}
 \email[ucotti@ifm.umich.mx]{}
\author{J. L. Diaz-Cruz$^2$}
 \email[ldiaz@sirio.ifuap.buap.mx]{}
\author{R. Gait\'an$^3$}
 \email[rgaitan@servidor.unam.mx]{}
\author{H. Gonzales$^4$}
\author{A. Hern\'andez-Galeana$^5$}
\affiliation{$^1$IFM-UMSNH, A.P. 2-82, 58041,
             Morelia, Mich. M\'exico.\\
         $^2$Instituto de F\'{\i}sica, BUAP, A.P. J-48, 72570 Puebla,
             Pue. M\'exico \\
         $^3$Centro de Investigaciones Te\'oricas, Facultad de Estudios
             Superiores-Cuautitlan,
             A.P. 142, Cuautitl\'an-Izcalli, Estado de M\'exico,
             C. P. 54770, M\'exico.  \\
         $^4$Departamento de F\'{\i}sica, Universidad
             Surcolombiana,
             A. A. 385, Neiva, Colombia.  \\
         $^5$Departamento de F\'{\i}sica, Escuela Superior de F\'{\i}sica
             y Matem\'aticas,
             Instituto Polit\'ecnico Nacional, U.P. Adolfo L\'opez
             Mateos, C. P. 07738, M\'exico.}

\date{\today}

\begin{abstract}
 We study the conditions under which flavor violation arises in
scalar-fermion interactions, as a result of the mixing phenomena
between the standard model and exotic fermions. Phenomenological
consequences are discussed within the specific context of a left-right
model where these additional fermions have mirror properties under
the new $SU(2)_R$ gauge group.
 Bounds on the parameters of the model are obtained from
LFV processes; these results are then used to study the LFV Higgs decays
($H\to \tau l_j$, $l_j=e,\mu$),  which reach branching
ratios that could be detected at future colliders.
\end{abstract}
\maketitle

                      \section{Introduction}
 The existence of additional fermions, and their possible mixing with
the known quarks and leptons of the Standard Model (SM), has been widely
studied in the literature.
 These additional fermions could behave like the SM ones, as in the case
of a fourth family (some times also called sequential fermions),
or have exotic properties (like vectorlike
or mirror fermions).
 In particular, mirror fermions are very interesting objects from a
phenomenological point of view, and their low-energy effects have
been studied too~\cite{Maalampi:1990va}, including a
discussion of limits on their masses~\cite{Groom:2000in}.
 More recently, models involving mirror fermions generations, i.e.,
fermions with ``mirror'' isospin charges, have been discussed within
a left-right (LR) symmetric context~\cite{Ceron:1998ey}.
 These models offer a possible solution to the strong CP problem, and have
also been discussed in the context of strongly-interacting
electroweak symmetry breaking scheemes~\cite{Triantaphyllou:1998ke,Lindner:1998er}.
 Applications to the search for new astrophysical phenomena have been
pursued too~\cite{Berezhiani:1995yi,Foot:1995pa}.

 In addition to direct production, these new fermions can manifest
themselves through their  mixing with the standard fermions, which
can affect  several aspects of flavor physics.
 For instance, new sources of flavor changing neutral process are
usually presented in these extensions of the SM,
which could lead to observable signatures and to seriously
constrain such models.
 Most applications have focused on the light fermions, and only
recently the possible implications for the top quark have been
studied~\cite{delAguila:1999ec,delAguila:1999ac}.
 Given the fact that coming stages of Tevatron (Run-II) will
produce about $10^4$ top pairs, it is interesting to study rare
decays of the top as possible tests of the
SM~\cite{Diaz-Cruz:1990ub,Diaz-Cruz:1999ab}.
 Furthermore, since Higgs bosons could also be
produced at significant rates at Tevatron, it is also important to
search for all the relevant Higgs signatures that could be detected
at Run-2; future colliders (LHC, VLHC, linear colliders) will have
the chance to extend further the search range.
 Possible rare flavor violating Higgs decay modes have been
overlooked thus far, and only recently their phenomenological
relevance has been discussed \cite{Diaz-Cruz:1999xe}.

 In this paper we analyze lepton-flavor-violating (LFV) signals in
the context of LR models with mirror fermions; we focus on
the mixing phenomena that arises from the interactions
between the standard and exotic fermions, and their
implications for the scalar Higgs sector.
 After discussing general conditions for the appearance of
mixing in the extended fermion sector (Sec.~\ref{fermmix}),
we present in Sec.~\ref{lrmm} the LR model with mirror fermions,
which is one of the simplest LR model that solves the strong CP problem.
 The model contains the gauge group
 $\rm SU(2)_L \times SU(2)_R \times U(1)$.
 We discuss both its fundamental aspects and
how the mixing between standard and mirror fermions can be
described within the formalism presented in Sec.~\ref{fermmix}.
 In this model we can identify two sources that can mediate
flavor-changing transitions, namely: the exchange of
family-violating neutral gauge bosons, and the exchange of Higgs
bosons with flavor-violating couplings. We shall prove that these
sources are described by independent parameters, and in order to
search for the optimal rates for the signals of our interest, we
shall assume that the bounds are saturated by the scalar exchange,
which could simple mean that the mass of the additional gauge
boson are much heavier than the SM-like Higgs boson of the model.

 In Sec.~\ref{lfv} we derive bounds on the parameters of the model
associated with LFV transitions, for which we shall include the
Higgs contributions to the transitions with strongest
bounds known, namely
$\mu \rightarrow e\gamma$,
$\tau \rightarrow \mu \gamma$,
$\tau \rightarrow e\gamma$,
$\mu \rightarrow eee$,
$\tau \rightarrow eee$, $\mu ee$, $\mu \mu e$, $\mu \mu \mu$,
and electron-muon conversion in nuclei.
 In Sec.~\ref{hlillj} we use the allowed range on the couplings
$H l_i l_j$, to evaluate the B.R. for the LFV decay modes
of the Higgs themselves, namely $H \rightarrow l_i l_j $,
as well as the prospects to detect this signal at the future colliders
(Tevatron, LHC and VLHC).
 Finally, Sec.~\ref{concl} contains our conclusions.

\section{Fermion mixing and flavor violation}
\label{fermmix}
 We shall present first a generic description of fermion mixing, when
two types of fermions are included: ordinary ($o$) and exotic ($e$).
 The ordinary fermions include the SM ones, whereas the exotic ones
include any new fermion with sequential, mirror or singlet properties,
beyond the SM.
 We shall assume that the SM Higgs gives masses to the SM fermions, and also
that there could be an unspecified Higgs sector responsible for the exotic
fermion masses, although in some cases gauge-invariant mass terms
could be allowed.

 To consider the mixing of fermions,  we shall follow
Ref.~\cite{Langacker:1988ur}, grouping all fermions of a given electric charge
$q$ and a helicity $a = L, R$ into $n_a + m_a$ vector column of $n_a$
ordinary ($o$) and $m_a$ exotic ($e$) gauge eigenstates, i.e.
$\psi_a^o =
 \left(
  \psi_o^o, \psi_e^o
 \right)_a^{\top}$.
 The VEV of the neutral SM scalar produces the SM fermion mass terms,
which together with the exotic mass and mixing matrices, will be
grouped as (working in the gauge eigenstate basis):
\begin{equation}
 {\cal L}_{\rm mass} =
 \overline{\psi^o_L} {\sf M} \psi^o_R + \rm h. c. .
\end{equation}
The mass matrix $\rm M$ takes the form
\begin{equation}
\label{massmatr}
 \sf M =
  \left(
   \begin{array}{cc}
    \mathsf K  &  \mu'   \\
    \mu  &  \mathsf K'
   \end{array}
  \right),
\end{equation}
where \textsf{K} denotes the SM fermion mass matrix
and \textsf{K'} corresponds
to the fermion mass matrices associated with the exotic sector,
while $\mu, \mu'$ correspond to the possible mixing
terms between ordinary and exotic fermions.

The form of the mass matrix will of course depend on the
type of exotic fermion to be considered. Some interesting
cases where one can discuss which terms arise from
Spontaneous Symmetry Breaking (SSB) and which
ones could be vectorlike mass terms, include the following:

\begin{itemize}

\item For a sequential fourth family,
  $\psi_{eL} \sim \bf{2}$  and
  $\psi_{eR} \sim \bf{1}$ under $SU(2)_L$, then
  all of $\mu,\mu', \sf K, K'$ arises from SSB,
  and become part of the full $4 \times 4$ mass matrix,
  which describes fermion masses and mixing.

\item For a trivial vector-like representation, where:
  $\psi_{eL} \sim \bf{1}$  and
  $\psi_{eR} \sim \bf{1}$ under $SU(2)_L$,
  the terms $\mu,\mu'$ will require SSB,
  but $\sf K'$ would be a gauge-invariant mass term.

\item For a vector-like representation under the SM gauge group,
 i.e. with  $\psi_{eL} \sim \bf{n}$  and
  $\psi_{eR} \sim \bf{\bar{n}}$  under $SU(2)_L$,
  then when  ($\bf{n} \neq 2$),
  $\mu=\mu'=0$ and $\sf K'$ will be a gauge-invariant mass term.

\item For a mirror family, i.e.
  $\psi_{eL} \sim \bf{1}$  and
  $\psi_{eR} \sim \bf{2}$, under $SU(2)_L$, then
  the mass terms $\mu,\mu'$ will be gauge-invariant
  terms, while $\sf K'$ will require SSB.

\item For the case when
     $\psi_{eL} \simeq \bf{1}$  and
  $\psi_{eR} \simeq \bf{1}$ under $SU(2)_L$, but now these fermions
  transform non-trivially under an additional gauge
  group, then $\sf K'$ will require SSB,
  and $\mu$ or $\mu'$ would be the allowed gauge-invariant
  mass terms. This is in fact the case that will be discussed in the
  next section, where the exotic fermions have mirror properties
  with respect to a new $SU(2)$ gauge group.

\end{itemize}

 The relation between the gauge eigenstates and the corresponding
light ($l$) and heavy ($h$) mass eigenstates
$\psi_a =
 \left(
  \psi_l, \psi_h
 \right)_a^{\top}$
is given by a the transformation
\begin{equation}
  \psi_a^o = {\sf U}_a  \psi_a
 \label{fermass}
\end{equation}
where
\begin{equation}
{\sf U}_a =
  \left(
   \begin{array}{cc}
   {\sf A}_a & {\sf E}_a  \\[2mm]
   {\sf F}_a & {\sf G}_a
  \end{array}
  \right).
\label{unit}
\end{equation}

 From the unitarity of $\sf U$
\begin{equation}
\left(
 {\sf U_a}{\sf U_a^{\dag}}
 \right) = \sf 1.
\end{equation}
it follows that the submatrix $\sf A_a$ is not unitary. The term
$\left(
  {\sf F^{\dag}} {\sf F}
  \right)_a,$
which is second order in the small mirror-standard fermion mixing,
will induce FC transitions in the light-light sector~\footnote{We shall %
identify the light fermions as the SM ones,
while the heavy ones ($\psi_h$) will consider the sector
beyond the SM, although this is not necessarily true, as a
consequence of the fermion mixing phenomena.}.

The diagonal mass matrix $\rm M_D$ can be  obtained through a biunitary
rotation acting on the $\rm L$ and $\rm R$ sectors, namely:
\begin{equation}
 {\sf M}_D = {\sf U}^\dag_L {\sf M} {\sf U}_R
 =
 \left(
  \begin{array}{cc}
   {\sf m}_l  &    \sf 0      \\[2ex]
     \sf 0    & {\sf m}_h
  \end{array}
 \right).
 \label{bota}
\end{equation}
where ${\sf m}_l, {\sf m}_h$ denote the light and heavy diagonal mass matrices,
respectively.
 We will also follow Ref.~\cite{Cotti:1997ev,Nardi:1993nq} to describe
the mixing
effects, though our work will concentrating mainly on the scalar
flavor-changing transitions.
 After substituting the expression for mass-eigenstates
eq.~(\ref{bota}), we can write the matrices \textsf{K}, \textsf{K'},
$\mu$ and $\mu'$ in terms of the mass-eigenvalues and the blocks
of matrix $\sf U$, as follows
\begin{equation}
 {\sf K}
 =
 \left(
  {\sf A}_L {\sf m}_l {\sf A}_R^\dag
  +
  {\sf E}_L {\sf m}_h {\sf E}_R^\dag
 \right),
\end{equation}
\begin{equation}
 {\sf K'}
 =
 \left(
  {\sf F}_L {\sf m}_l {\sf F}_R^\dag
  +
  {\sf G}_L {\sf m}_h {\sf G}_R^\dag
 \right),
\end{equation}
\begin{equation}
 \mu
 =
 {\sf F}_L {\sf m}_l  A_R^\dag +
 {\sf G}_L    {\sf m}_h E_R^\dag,
\end{equation}
\begin{equation}
 \mu' = \sf A_{\rm L} m_l  F_{\rm R}^\dag +
            E_{\rm L} m'_h G_{\rm R}^\dag .
\end{equation}

When one describes the scalar-fermion couplings within some specific
Higgs sector, it turns out that those couplings will not be diagonal
in general, thus new FCNC phenomena associated  will be predicted in
such model. It is relevant to mention here, that most analysis of
FCNC/LFV process in extended gauge models, have concentrated on
the flavor changing vertices of the additional neutral gauge bosons, but
the Higgs sector has been mostly overlooked.
 As it will be shown in the next sections, FCNC phenomena
associated with the scalar sector can produce new
discovery signals, consistent with the current bounds on FCNC/LFV
transitions. The resulting rates could be detected at coming stages
of colliders (RUN-2 of Tevatron) or in the future ones (LHC and
VLHC). The specific Higgs sector that we will work with,
corresponds to the minimal LR model with mirror fermions. This model
and the resulting phenomenological constraints, will be discussed
in the coming sections.

\section{Fermion mixing in a Left-Right model with mirror fermions}
\label{lrmm}

A first speculation on the existence of mirror fermions appeared
in the classical paper on parity violation that led to the V-A
interactions models~\cite{Lee:1956qn}.
 On the other hand, the left-right symmetric
models renders the baryon-lepton number symmetry $\rm U(1)_{B-L}$
more natural by gauging it, and it has also been proposed as a
solution to the strong CP problem when accompanied by the
introduction of mirror fermions~\cite{Barr:1991qx,Babu:1990rb}.
 Such an extended gauge
group with additional fermions, apart from fitting nicely into
unification schemata, restores the left-right symmetry missing in
the Standard Model (SM) in a manner that goes beyond the simplest
left-right symmetry models.

 In the left-right model with mirror fermions (LRMM)~\cite{Ceron:1998ey}, the
right-handed (left-handed) components of mirror fermions transform as
doublets (singlets) under the new $\rm SU(2)_R$. The SM fermions are singlets
under the $\rm SU(2)_R$, whereas the right-handed mirror are also singlets
under $\rm SU(2)_L$. Mirror and SM fermions will share hypercharge and
color interactions. Thus, the first family of leptons and quarks will be
written as follows:

\begin{equation}
\begin{array}{lr@{\extracolsep{8ex}}l@{\extracolsep{0ex}}r}
  l^o_{e L} =
  \left(
    \begin{array}{l}
     \rm \nu^o_e \\
     \rm e^o
    \end{array}
  \right)_L,
  &
  e^o_R,
  &
  {l'}^o_{e R}
 =
 \left(
  \begin{array}{l}
   {\nu'}^o_e  \\
   {e'}^o
  \end{array}
 \right)_R,
 &
 {e'}^o_L
\\[3ex]
 q^o_L =
 \left(
  \begin{array}{l}
   u^o \\
   d^o
  \end{array}
 \right)_L,
  &
  u^o_R,
  d^o_R,
  &
  {q'}^o_R =
 \left(
  \begin{array}{l}
   {u'}^o\\
   {d'}^o
  \end{array}
 \right)_R,
 &
 {u'}^o_L,
 {d'}^o_L
\end{array}
\end{equation}

The superscript ($o$) denote weak eigenstates, and the primes
will be associated with the mirror particles.
Because the model does not contain left-handed mirror
neutrinos, they have to be massless. Extensions of the model to include
neutrino masses has been considered too \cite{Foot:1995pa}, but it is
beyond the scope of our present work.

subsection{Symmetry Breaking}
\label{symm-break}
 The symmetry breaking is realized by including two Higgs doublets, the SM
one ($\phi$) and its mirror partner ($\phi'$). The potential of the
model can be written in a such a way that the vacuum expectation values
(VEV's) of the Higgs field are
\begin{equation}
 \langle \phi \rangle
 =
 \frac{1}{\sqrt{2}}
 \left(
  \begin{array}{l}
   0\\
   v
  \end{array}
 \right),
 \hspace{10mm}
 \langle \phi' \rangle
 =
 \frac{1}{\sqrt{2}}
 \left(
  \begin{array}{l}
   0 \\
   v'
  \end{array}
 \right).
\label{vac}
\end{equation}

The most general potential that develops this pattern of VEV's is:
\begin{equation}
 V
 =
 -\left(
  \mu_1^2 \phi^\dag \phi + \mu_2^2 {\phi'}^\dag \phi'
 \right) +
 \frac{\lambda_1}{2}
 \left(
  \left(
   \phi^\dag \phi
  \right)^1 +
  \left(
   {\phi'}^\dag \phi'
  \right)^2
 \right) +
 \lambda_2
 \left(
  \phi^\dag \phi
 \right)
 \left(
  \phi^\dag \phi
 \right).
\end{equation}

The terms with $\mu_1, \mu_2$ are included  so that the
parity symmetry (P) is broken softly, i.e. only through the
dimension-two mass terms of the Higgs potential.

The neutral Higgs boson squared mass matrix that follows from
this potential is:
\begin{equation}
 \rm {\sf M}^2_{H^0}
 =
 \left(
  \begin{array}{cc}
   2 \lambda_1 v^2   &  2 \lambda_2 v v' \\
   2 \lambda_2 v v'  &  2 \lambda_1 {v'}^2
  \end{array}
 \right) .
\end{equation}

Diagonalization of the higgs-boson squared mass matrix is
straightforward using a real basis. Out of the eight scalar
degrees of freedom associated with two complex doublets, six
become the Goldstone bosons required to give mass to $W^\pm$,
${W'}^\pm$, $Z$ and $Z'$. Thus only two neutral Higgs bosons remain;
the neutral physical states are:
\begin{eqnarray}
 H
 =
 \sqrt{2}
 \left(
  \left(
   \Re e \phi^o -v
  \right)
  \cos\alpha +
  \left(
   \Re e {\phi'}^o -v'
  \right)
 \sin\alpha
 \right), \\[3mm]
 H'
 =
 \sqrt{2}
 \left(
 -\left(
  \Re e \phi^o -v
 \right)
 \sin\alpha +
 \left(
  \Re e {\phi'}^o -v'
 \right) \cos\alpha
\right),
\end{eqnarray}
where $\alpha$ denotes the neutral Higgs mixing angle.

The mass matrix for the gauge bosons is obtained from the scalar
Lagrangian
\begin{equation}
 \rm {\cal L}_{sca}
 =
 \left(
  D^\mu \phi
 \right)^\dag
 \left(
  D_\mu \phi
 \right)
 +
 \left(
  {D'}^\mu \phi'
 \right)^\dag
 \left(
  D'_\mu \phi'
 \right)
\label{boson}
\end{equation}
where $D_\mu$ denotes the covariant derivative associated with the
SM, and $D'_\mu$ is the one associated with the mirror sector.

 After substituting the vev's~(\ref{vac}) in the
Lagrangian~(\ref{boson}) we obtain the expressions for the mass matrices.
 The mass matrix for the charged gauge bosons is already diagonal,
with mass eigenvalues: $ M_W = \frac{1}{2} vg_2$ and $M_{W'}
= \frac{1}{2} v'g'_2$, where $g_2$ and $g'_2$ are the coupling constant
associated with the $\rm SU(2)_L$ and $\rm SU(2)_R$ gauge group,
respectively.
 The mass matrix for the neutral gauge bosons is not diagonal,
and is given by
\begin{equation}
 \frac{1}{4}
 \left(
  \begin{array}{ccc}
   g^2_2         & 0             &  -g_2  g_1 v^2 \\
    0            & {g'_2}^2 {v'}^2 &  -g'_2 g_1 {v'}^2\\
   -g_2 g_1 v^2  & g'_2 g_1 {v'}^2 &  g_1^2 \left( v^2 + {v'}^2
  \right)
 \end{array}
\right) .
\end{equation}

However this matrix can be diagonalized by an orthogonal transformation
$\sf R$, which relates the weak and mass eigenstates,
namely~\cite{Ceron:1998ey}
\begin{equation}
 \sf R = \left(
 \begin{array}{ccc}
  c_{\theta_w}  c_{\Theta} & c_{\theta_w}  s_{\Theta} &
 s_{\theta_w}\\
 -\frac{1}{c_{\theta_w}} \left( s_{\Theta} r_{\theta_w} +
 \frac{g_2}{g'_2} c_{\Theta} s^2_{\theta_w} \right) &
 \frac{1}{c_{\theta_w}} \left( c_{\Theta} r_{\theta_w} -
 \frac{g_2}{g'_2} s_{\Theta} s^2_{\theta_w} \right) &
\frac{ g_2}
 {g'_2} s_{\theta_w} \\
 t_{\theta_w} \left( \frac{ g_2}{g'_2} s_{\Theta} -
 r_{\theta_w} c_{\Theta} \right) & - t_{\theta_w} \left( \frac{
 g_2}{g'_2} c_{\Theta} + r_{\theta_w} s_{\Theta} \right) &
 r_{\theta_w}
 \end{array}
 \right),
 \label{rot}
\end{equation}
with $\theta_w$ and $\Theta$ denoting the rotations angles of the
neutral gauge bosons. In
eq.~(\ref{rot}), $r_{\theta_w} \equiv \sqrt{c^2_{\theta_w} -
                 \frac{g^2_2}{{g'}^2_2} s^2_{\theta_w}}$.
One has one massless state (the photon) and two massive states, with
eigenvalues given by
\begin{equation}
  M_{Z, Z'}
 =
 \frac{1}{2}
 \left(
  v^2
  \left(
   g^{2}_{2} + g_{1}^{2}
  \right)
  +
  {v'}^2
  \left(
   {g'}^2_2 + g_{1}^{2}
  \right)
 \right)
 \mp
 \frac{1}{2}
 \sqrt{
  \left(
   v^2
   \left(
    g^{2}_{2} + g_{1}^{2}
   \right)
   +
   {v'}^2
   \left(
    {g'}^2 + g_1^2
   \right)
  \right)^2
  -
  4 v^2 {v'}^2
  \left(
   g^2_2 g_1^2 + g^2_2 {g'}_2^2 + {g'}_2^2 g_{1}^{2}
  \right)
 },
\end{equation}
where $ g_1$ is the coupling constant of the $U(1)$ gauge group.

In order to find the couplings of the Higgs fields
with the neutral gauge boson,
we expand the expressions for $D_{\mu}$ and
$D'_\mu$, substituting the physical states
into eq.~(\ref{boson}). We get the following expression for
the $ZZH/H'$ interactions:
\begin{eqnarray}
  {\cal L}_{ZZH}
 =
 \sqrt{2}
 \left(
  g_2 M_W X
  \left(
   \Theta, \theta_w
  \right)
  \cos\alpha
  +
  g'_2 M_{W'} Y
  \left(
   \Theta, \theta_w
  \right)
  \sin\alpha
  \right)
  H Z_\mu Z^\mu \nonumber\\
  +
  \sqrt{2}
  \left(
   -g_2 M_W X
   \left(
    \Theta, \theta_w
   \right)
   \sin\alpha
   +
   g'_2 M_{W'} Y
   \left(
    \Theta, \theta_w
   \right)
   \cos\alpha
  \right)
  H' Z_\mu Z^\mu
\end{eqnarray}
where
\begin{equation}
  X
 \left(
  \Theta, \theta_w
 \right)
 =
 \left(
  c_{\theta_w} c_\Theta - \frac{g_1}{g_2} t_{\theta_w}
  \left(
   \frac{g_2}{g'_2} s_\Theta - r_{\theta_w} c_\Theta
  \right)
 \right)^2,
\end{equation}

\begin{equation}
  Y
 \left(\Theta, \theta_w
 \right)
 =
 \left(
  -\frac{1}{c_{\theta_w}}
  \left(
   s_\Theta r_{\theta_w} + \frac{g_2}{g'_2} c_\Theta s^2_{\theta_w}
  \right)
  -
  \frac{g_1}{g_2} t_{\theta_w}
  \left(
   \frac{g_2}{g'_2} s_\Theta - r_{\theta_w} c_\Theta
  \right)
 \right)^2
\end{equation}

On the other hand, the expressions for the Higgs-charged gauge boson
interactions ($WWH/H'$) are given by:
\begin{eqnarray}
  {\cal L}_{W W H}
 =
 \sqrt{2} g_2 M_W g^{\mu \nu}
 \left(
  H \cos\alpha - H' \sin\alpha
 \right)
 W^-_\mu W^+_\nu
\end{eqnarray}

    \subsection{Yukawa Lagrangian and fermion mixing}
\label{yuklagr}

The renormalizable and gauge invariant interactions of the scalar
doublets $\phi$ and $\phi'$ with the leptons are described by the
Yukawa Lagrangian, which takes the form
 \begin{equation}
   {\cal L}^l_Y = 
   \lambda_{i j} \overline{l}^o_{i L} \phi e^o_{j R} +
   \lambda'_{i j} \overline{l'}^o_{i R} \phi' {e'}^o_{j L} +
   \mu_{i j} \overline{e'}^o_{i L} e^o_{j R} + h. c.
 \end{equation}
where $ i, j = 1, 2, 3$ and $\lambda_{ i j}$,
$\hat{\lambda}_{ i j}$, and $\mu_{ i j}$ are (unknown)
matrices.

For the quarks fields, the corresponding Yukawa terms are written as
 \begin{equation}
  {\cal L}^q_Y
  =
  \lambda^d_{ij}  \overline{Q}^o_{iL} \phi d^o_{jR} +
  \lambda^u_{ij}  \overline{Q}^o_{iL} \tilde{\phi} u^o_{jR} +
  {\lambda'}^d_{ij} \overline{Q'}^o_{iR} \phi' {d'}^o_{jL} +
  {\lambda'}^u_{ij} \overline{Q'}^o_{iR} \tilde{\phi'} {u'}^o_{jL} +
  \mu^d_{ij}      \overline{d'}^o_{iL} d^o_{jR} +
  \mu^u_{ij}      \overline{u'}^o_{iL} u^o_{jR} + h.c.,
 \end{equation}
the conjugate fields $\tilde{\phi}$ ($\tilde{\phi'}$) are
obtained as $\tilde{\phi} =  i \tau_2 \phi^*$.

The VEV's of the neutral scalars produce the fermion mass terms, which
in the gauge eigenstate basis read
 \begin{equation}
  {\cal L}_{\rm mass}
  =
  \overline{\psi^o_L} {\sf M} \psi^o_R + h.c. .
 \end{equation}
For the lepton sector, the non-diagonal mass matrix $\sf M$,
takes the form
\begin{equation}
\sf M
 =
 \left(
  \begin{array}{cc}
   \sf K   & \sf 0       \\
   \sf \mu & \sf K'
  \end{array}
 \right),
\end{equation}
where $\sf K = \frac{1}{2} {\lambda} v$ and $\sf K' =
\frac{1}{2} \lambda' v'$ correspond to the $3 \times 3$
matrices generated from the symmetry breaking VEV's; $\mu$
corresponds to the gauge invariant $3 \times 3$ mixing terms between
ordinary and mirror fermions singlets.
 Thus, we have  $\mu'=0$ in Eq.~(\ref{massmatr}), and then:
\begin{equation}
 {\sf M}_D = {\sf U}^\dag_L {\sf M} {\sf U}_R .
\end{equation}

 With the help of the relations (7-10), and working within the
Higgs mass-eigenstate basis, the tree-level interactions of the
neutral Higgs bosons H and H' with the light fermions are given by
\begin{eqnarray}
 {\cal L}_Y^l
 =
 \frac{g_2}{2 \sqrt{2}}
 \overline{f}_L {\sf A}^\dag_L  {\sf A}_L \frac{m_l}{M_W} f_{R}
 \left(
  H \cos\alpha - H' \sin\alpha
 \right)
 +
 \frac{g'_2}{\sqrt{2}}
 \overline{f}_L \frac{m_l}{M_{W'}}
 {\sf F}^\dag_R {\sf F}_R f_R
 \left(
  H \sin\alpha + H' \cos\alpha
 \right) + h.c. .
\end{eqnarray}
One can see that the couplings are not diagonal in general, thus new
phenomena associated with FCNC will be present in this model. The
resulting phenomenological constraints and predictions will be discussed
in the coming sections.

Finally, once we have obtained the quark and lepton mass eigenstates,
their gauge interaction can be obtained from the
Lagrangian
 \begin{equation}
  {\cal L}^{\rm int}
  =
  \overline{\psi}  i \gamma^\mu D^\mu \psi +
  \overline{\psi'} i \gamma^\mu D'_\mu \psi'
  \label{inter}
 \end{equation}
where $\psi$, $\psi'$ denote the standard and mirror fermions,
respectively.

The neutral current term for the multiplet $\psi$ of a given
electric charge, including the contribution of the neutral gauge
boson mixing, can be written as follows
 \begin{equation}
  -{\cal L}^{\rm nc}
  =
  \sum_{a =  L, R} \overline{\psi}_a^o \gamma^\mu
  \left(
   g_2 {\sf T}_{3 a}, g'_2 {\sf T}'_{3 a}, g_1 \frac{{\sf Y}_a}{2}
  \right)
  \psi_a^o
  \left(
   \begin{array}{l}
    W^3\\
    {W'}^3\\
    B
   \end{array}
  \right)_\mu
 \end{equation}

Using Eqs.~(\ref{fermass},\ref{unit}),
one arrives to the following expression
in terms of the mass eigenstates:
\begin{equation}
 -{\cal L}^{\rm nc}
 =
 \sum_{a = L,R} \overline{\psi}_a \gamma^\mu {\sf U}_a^\dag
 \left(
  g_2 {\sf T}_{3a}, g'_2 {\sf T'}_{3a}, g_1 \frac{{\sf Y}_a}{2}
 \right)
 {\sf U}_a \psi_a
 \left(
  \begin{array}{l}
   Z\\
   Z'\\
   A
  \end{array}
 \right)_\mu
\end{equation}
where ${\sf T}_{3a}$, ${\sf T'}_{3a}$, and Y are the generators of the
$\rm SU(2)_L$, $\rm SU(2)_R$, and U(1), respectively.

As we mentioned before, in the coming analysis we shall concentrate on the
scalar mediated flavour-violating transitions, assuming that they
saturate current bounds on FCNC/LFV transitions, which will also
simplify the analysis.

\section{Constraints from LFV transitions mediated by the Higgs boson}
\label{lfv}

 In the following, we shall discuss the constraints that low-energy
data imposes on the parameters of the LFV higgs lagrangian.
 After neglecting the lighter lepton masses in each vertex,
we arrive to the following expressions for the LFV Higgs interactions:
\begin{equation}
{(H l_i l_j)}:  \frac{g_2 m_j}{2m_W} \eta_{ij} P_R \\
\end{equation}
with $m_j > m_i$ and $P_R=(1+\gamma_5)/2$.
For the diagonal interactions we obtain,
\begin{equation}
{(H l_i l_i)}: \frac{g_2 m_i}{2m_W} \eta_{ii}
\end{equation}
In eqs. (34) and (35),
$\eta_{ij}= (A^\dagger_L A_L)_{ij} \cos\alpha$.
In fact, we shall see that the resulting bounds
can be expressed in terms of the
parameters $\eta_{ij}/m^2_H$.

\subsection{The decays $l_i \to l_j l_k l_k$}
\label{liljlk}
The flavor-changing interactions of the scalars
$H$ and $H'$ will mediate the lepton number violation
decays $l_i \to l_j l_k l_k$. By assuming  $m_{H'} >> m_H$,
one can neglect the contribution of the mirror Higgs. Then,
using the interactions contained in eqs.(35-36) we can derive
the expressions for the decay width of these
processes. The result is given by:

\begin{equation}
\begin{array}{r}
\label{roa}
\Gamma =
\frac{m_i^5}{3072 \pi^3 M_H^4}
\left(
 \left(
  \left|
   g^{ik}_V
  \right|^2
  +
  \left|
   g^{ik}_A
  \right|^2
  \right)
   \left(
    \left|
     g^{jk}_V
    \right|^2
    +
    \left|
     g^{jk}_A
    \right|^2
   \right)
   +
   \left(
   \left|
    g^{ij}_V
   \right|^2
   +
   \left|
   g^{ij}_A
  \right|^2
  \right)
   \left(
    \left|
     g^{kk}_V
    \right|^2
   \right)
   +
  \right.
 \\[3ex]
 \left.
 \frac{1}{2}
  \left(
   g^{ik}_V g^{jk}_V g^{ij}_V g^{kk}_V
   +
   g^{ik}_V g^{kk}_V g^{jk}_A g^{ij}_A
   +
   g^{jk}_V g^{kk}_V g^{ik}_A g^{ij}_A
   +
   g^{ij}_V g^{kk}_V g^{ik}_A g^{jk}_A
  \right)
 \right).
\end{array}
\end{equation}

This amplitude includes two possible Feynman graphs, one with two
LFV vertices and one with a flavor conserving vertex.
The explicit expressions for the scalar $g_V^{ij}$
and pseudoscalar $g_A^{ij}$ couplings associated with
the lepton transitions $ij$ (with $i\neq j$) mediated by
the scalar $H$, are given by:
\begin{equation}
 g_V^{ij}
 =
 g_A^{ij}
  =
 \frac{g_2}{4 M_W} \eta_{ij} m_j
\end{equation}
whereas
\begin{equation}
 g_V^{kk}
 =
 \frac{g_2}{2 M_W} \eta_{kk} m_j,
 \hspace{6ex}
 g_A^{kk}
 =
 0 .
\end{equation}

In deriving Eq.~(\ref{roa}), we have neglected
the lepton masses of the final states.
On the other hand, we have neglected the $Z$-fermions interactions
because they are diagonal in this approximation.

Now, we examine the three-body decays $\rm \mu \rightarrow eee$, $\rm \tau
\rightarrow eee$, $\rm \tau \rightarrow \mu \mu \mu$, $\rm \tau
\rightarrow \mu \mu e$, $\rm \tau \rightarrow ee \mu$, $\rm \tau
\rightarrow e \mu e$, and $\rm \tau \rightarrow e \mu \mu$, which can be
induced by the $LFV$ couplings. Using the expressions for the partial
withs and comparing with the experimental limits, we derive bounds on
the flavor changing couplings $\eta_{\rm i j}$ as function of $M_{H}$.
The results are:
\begin{enumerate}
\item
$\mu \rightarrow eee$
\begin{equation}
{\eta^2_{ee}} {\eta^2_{\mu e}} < 5.5 \times 10^{-3} M^4_{H}
{GeV}^{-4}
\end{equation}
\item
$\tau \rightarrow eee$
\begin{equation}
{\eta^2_{ee}} {\eta^2_{\tau e}} < 3.2 \times 10^{2} M^4_{H}
{GeV}^{-4}
\end{equation}
\item
$\tau \rightarrow \mu \mu \mu$
\begin{equation}
{\eta^2_{\mu \mu}} {\eta^2_{\tau \mu}} < 4.8 \times 10^{-3} M^4_{H}
{GeV}^{-4}
\end{equation}
\item
$\tau \rightarrow \mu \mu e$
\begin{equation}
{\eta^2_{\tau \mu}} {\eta^2_{\mu e}} < 7.6 \times 10^{-3} M^4_{H}
{GeV}^{-4}
\end{equation}
\item
$\tau \rightarrow ee \mu$
\begin{equation}
{\eta^2_{\tau e}} {\eta^2_{\mu e}} < 7.6 \times 10^{-3} M^4_{H}
{GeV}^{-4}
\end{equation}
\item
$\tau \rightarrow e \mu e$
\begin{equation}
{\eta^2_{ee}} {\eta^2_{\tau \mu}} + 2 \times 10^4 {\eta^2_{\tau e}}
{\eta^2_{\mu e}} + 2 \times 10^2 {\eta_{ee}} {\eta_{\tau e}} {\eta_{\mu
e}} {\eta_{\tau \mu}} < 2 \times 10^2 M^4_{H} {GeV}^{-4}
\end{equation}
\item
$\tau \rightarrow e \mu \mu$
\begin{equation}
{\eta^2_{\tau e}} {\eta^2_{\mu \mu}} + \frac{1}{2} {\eta^2_{\tau \mu}}
{\eta^2_{\mu e}} + \frac{1}{2} {\eta_{\tau e}} {\eta_{\mu \mu}} {\eta_{\mu
e}} {\eta_{\tau \mu}} < 4.6 \times 10^{-3} M^4_{H} {GeV}^{-4}
\end{equation}
\end{enumerate}

For instance, taking $M_{H}$ = 130 $GeV$ gives the following
bounds:
${\eta^2_{ee}} {\eta^2_{\mu e}} < 1.6 \times 10^6$, ${\eta^2_{\mu \mu}}
{\eta^2_{\mu \tau}} < 1.4 \times 10^6$, ${\eta^2_{\tau \mu}}
{\eta^2_{\mu e}} < 2.2 \times 10^6$, and ${\eta^2_{\tau e}} {\eta^2_{\mu
e}} < 2.2 \times 10^6$.
These results show that the LFV Higgs couplings are essentially
unconstrained by low-energy phenomenology. However, since they
appear as products of fermion and Higgs mixing angles, it is
natural to expect them to be naturally of order 1; moreover,
for the application to LFV Higgs decay, we shall take them
conservatively to be of order 0.1.


\subsection{Radiative decays}
\label{meg}

Here we analyze the lepton flavor violation processes $ \mu
\rightarrow e \gamma$, $\tau \rightarrow e \gamma$ and $\tau
\rightarrow \mu \gamma$, arising in the model as a consequence of
the existence of the gauge invariant mixing terms $ \mu_{\rm i j}
\overline{\hat{\rm e}^o_{\rm i L}} {\rm e}^o_{\rm j R}$ + h.c of
ordinary leptons with mirror counterparts. We discuss here the main
details for the process $\mu \rightarrow e \gamma$, with the other
processes being computed in an analogous way.

The lowest order contribution to the $\mu \rightarrow e \gamma$ decay
mediated by the neutral scalar fields comes from the Feynman diagrams
where the photon is radiated from an internal lepton line.
As it is known the corresponding amplitude is proportional to the
operator
$\overline{\rm u(p_2)}\sigma^{\mu \nu} q_{\nu} \epsilon_{\mu} \rm u(p_1)$,
where $q =p_1 - p_2$ and $\epsilon_{\mu}$ is the polarization of the photon;
an helicity flip is therefore involved.
The amplitudes obtained by computing the one loop diagrams are
\begin{equation}
\label{olga1}
  i {\rm m_{A}}  =
  C_{A} \bar{e} i
    \frac{\sigma^{\mu \nu} q_{\nu} \epsilon_{\mu}}
     {m_{\mu} + m_e}
    \frac{1 - \gamma_5}{2} \mu
\end{equation}
and
\begin{equation}
  i {\rm m_{B}}  =
\label{olga2}
   C_{B} \bar{e} i
   \frac{\sigma^{\mu \nu} q_{\nu} \epsilon_{\mu}}
    {m_{\mu} + m_e}
    \frac{1 + \gamma_5}{2} \mu,
\end{equation}
corresponding to the contributions to $\mu_{\rm L} \rightarrow
e_{\rm R} \gamma$ and $\mu_{\rm R} \rightarrow e_{L} \gamma$
respectively.

From the unitary of $\sf U$ it follows that
\begin{equation}
{\sf A^{\dag}} {\sf A} =
 \sf 1 - {\sf F^{\dag}} {\sf F}
\end{equation}
 Hence, the matrix $\sf A$ describing the mixing with the ordinary
charged leptons is non-unitary by small terms quadratic in the
ordinary-exotic charged lepton mixing present in $\sf F$. In
our calculation we keep only the lowest order in the non-diagonal terms.
So, the leading contribution to the coefficients $C_{A}$ and $C_{B}$,
in the limit $\alpha \ll 1$, $M_{W} \ll M_{\hat{W}}$, and $M_{H} \ll M_{H'}$
is given by
\begin{equation}
 C_A
 =
 \frac{e}{16 \pi^2}
 \frac{g^2}{4 M_H^2 M_W^2}
 \left(
  m_\mu
  +
  m_e
 \right)
  m_\mu^2 m_e
  \left[
   \ln \frac{M_H^2}{ m_\mu^2} - \frac{4}{3}
  \right]
  \left(
  {\sf A}^\dag_L {\sf A}_L
 \right)_{12}
\end{equation}
and
\begin{equation}
C_B  =
 \frac{e}{16 \pi^2}
 \frac{g^2}{4 M_H^2 M_W^2}
  \left(
   m_\mu + m_e
    \right)
     m_\tau^{3}
     \left[
      \ln \frac{M_H^2}{m_\mu^2} - \frac{4}{3}
       \right]
        \left(
         {\sf A}^\dag_L {\sf A}_L
          \right)_{12}
\end{equation}

After comparing Eq.~(\ref{olga1}) and Eq.~(\ref{olga2}) with the general expressions from
Ref.~\cite{Lee:1977ti} for a
process with a real photon $f_1 \longrightarrow f_2 + \gamma$:
\begin{eqnarray}
 i{\cal M}
  \left(
   f_1(p_1) \rightarrow f_2(p_2) + \gamma(q)
  \right)
  &
  =
  &
  \bar{u}_2(p_2) i
  \frac{\sigma^{\mu \nu} q_\nu \epsilon_\mu}{m_1 + m_2}
  \left(
   F(0)^V_{21} + F(0)^A_{21} \gamma_5
  \right)
  u_1(p_1), \\[3mm]
{\Gamma (f_1 \rightarrow f_2 + \gamma)} &=&
 \frac{m_1}{8 \pi}
 \left(
  1 - \frac{m_2}{m_1}
 \right)^2
 \left(
  1 - \frac{m_2^2}{m_1^2}
 \right)
 \left[
  \left|
   F(0)^V_{21} \right|^2 +
  \left|
   F(0)^A_{21}
  \right|^2
 \right]
\end{eqnarray}

We find in our case:
\begin{eqnarray}
 F(0)^V_{21} &=&
    \frac{1}{2}
     \left(
      C_{A} + C_{B}
      \right), \\[3mm]
 F(0)^A_{21} &=&
      \frac{1}{2}
       \left(
        C_{B} - C_{A}
        \right)
\end{eqnarray}
and in the limit $m_{e} \ll m_{\mu} \ll m_{\tau}$ we get
\begin{eqnarray}
\label{Bmeg1}
 {\Gamma (l_i \rightarrow  l_j+\gamma)} =
   \frac{\alpha_{e.m.}}{512 \pi^{4}}
   \left(
    G_F m_{l_i}^{2}
    \right)^2
     \frac{m_{l_i}^{5}}{M_{H}^{4}}
     \left[
      \ln{\frac{M_{H}^{2}}{m_{l_i}^{2}}} - \frac{4}{3} \right]^2
       \left|
       \left(
        {\sf A}^\dag_L {\sf A}_L
   \right)_{ij}
   \right|^2
\end{eqnarray}

The bounds on the flavor-changing couplings which are obtained
by comparing these radiative decays with current limits,
are, respectively,
$\eta^2_{e \mu} < 1.9 \times 10^2$, $\eta^2_{\tau e} < 9.2 \times
10^3$, and $\eta^2_{\tau \mu} < 3.9 \times 10^3$, where we have taken
$M_{H}$ = 130 $GeV$ and $\eta_{ii}\simeq 1$.

\subsection{Bounds from electron-muon conversion}
\label{emu}
 Neutrinoless $\mu^{-} (A, Z) \rightarrow e^{-} (A, Z)$ conversion in
muonic atoms with mass number $A$ and atomic number $Z$ offer
another sensitive test of
$LFV$~\cite{Altarelli:1977zq,Chang:1994hz,Bernabeu:1993ta}.
 The experimental bound on the branching ratio for $\mu$ - $e$
conversion in titanium gives a constraint on possible
violation of the muon and electron numbers.
 In this section we obtain a bound on the parameter
$\eta_{\mu e}$ implied by the limit for the $LFV$ reaction,
as mediated by the light Higgs of the model.
 Using the results of Ref.~\cite{Ng:1994ey}, we can write:
\begin{eqnarray}
{\left(
 \left|
  g^{\mu e}_{V}
  \right|^2 +
   \left|
   g^{\mu e}_{A}
    \right|^2
     \right)^{\frac{1}{2}}} \leq
      \frac{2.4}{\sqrt{2}} \times 10^{-7}
        \left(
         \frac{{0.5} {GeV}}{\overline{m_{N}}}
         \right)
           \left(
            \frac{R}{10^{-16}}
             \right)^{\frac{1}{2}}
              \frac{M^2_{H}}{M^2_{W}}
\end{eqnarray}
where
\begin{eqnarray*}
\overline{m_{N}} = \frac{1}{2}
\left(
 m_{u} + m_{d}
 \right) = 5 {MeV}
\end{eqnarray*}
is the current quark mass and $R \simeq 10^{-16}$ is the
present sensitivity. Taking the values of $g^{\mu e}_{V}$ and
$g^{\mu e}_{A}$ of our model, we obtain:
\begin{equation}
\eta^2_{\mu e} \leq 8.6 \times 10^{-6} M^2_{H} {GeV}^{-2}
\end{equation}
Taking $M_{H}$ = 130 $GeV$, gives the result
$\eta^2_{\mu e} \leq 1.5 \times 10^{-1}$,
which is in fact the stronger bound.

\section{Detection of the LFV Higgs decays ($H\to l_i l_j$) }
\label{hlillj}
 The search for the Higgs boson is one of the main goals of
Tevatron RUN-2 and future colliders~\cite{Carena:2000yx}. Although
the most conservative search strategy uses the theoretical
expectations coming from the minimal standard model (SM), it is
certainly worthwhile to look for other signals arising from
physics beyond the SM. In this regard, it has been recognized
recently that the Higgs sector of several well-motivated  models
can predict  lepton flavor violating (LFV) Higgs decays with
sizable values~\cite{Diaz-Cruz:1999xe} without being in conflict with
low-energy phenomenology, and at rates that may be detectable at future
colliders. The case of the generic two-higgs doublet
model  III, has been further studied for linear and hadron
colliders~\cite{Sher:2000uq,Han:2000jz}, with the conclusion that
it is possible to detect the LFV HIggs signal, and this also
holds for the effective lagrangian extension of the SM.
 Within the minimal SUGRA-MSSM  and the SM
with massive neutrinos, these decays are found to have negligible
rates. Whereas in models with heavy majorana  neutrinos, the LFV
Higgs decays are induced at one-loop level and the
Branching  ratio (B.r.) can reach
values of order $10^{-3}$~\cite{Pilaftsis:1992st}.

In this paper we also study  the detectability of the
resulting LFV Higgs decays that arise in the LRMM.
In addition to considering the reach of Tevatron and LHC, we also
estimate the bounds on the LFV Higgs couplings that
could be obtained at the VLHC.

\subsection{The Branching ratios for LFV Higgs decays}
 One additional implication of these LFV couplings is the
possibility to observe the LFV Higgs decays $h^0 \to l^+_i l^-_j$,
whose decay width is given by:
\begin{equation}
 \Gamma(h \to l_i l_j) =  \frac{g^2_2 m^2_j m_H }{ 12\pi m^2_W}
|\eta_{ij}|^2
\end{equation}

\begin{figure}
 \includegraphics[scale = .55, angle = -90]{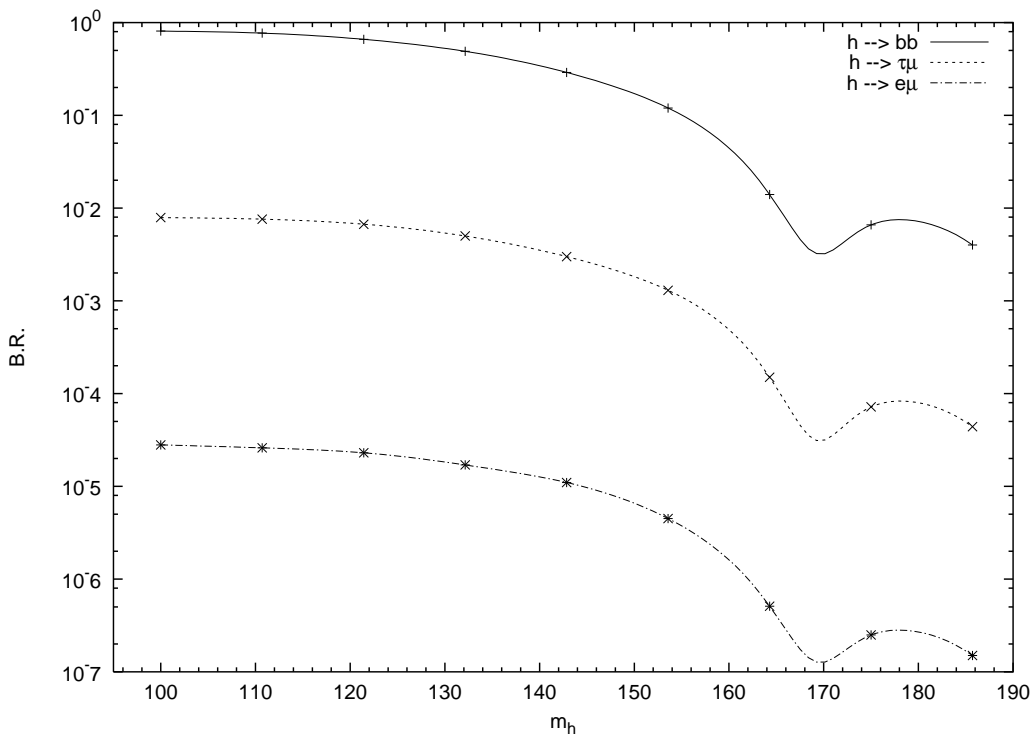}%
 \caption{ Branching ratios of the Higgs boson, with $\kappa_{\tau\mu}=0.1$.
  Solid ($b\bar{b}$), Dashes ($\tau-\mu$) and dot-dash ($e-\mu$).}
 \label{fig1}
\end{figure}

For current favored range of Higgs masses
($115 < m_h < 200$ GeV), one of the dominant decay modes
of the Higgs boson is into $h \to b \bar b$,
with the corresponding coupling being proportional to
$\eta_{bb}$, and this will introduce a complicated expression
for the Higgs total width. In order to handle such multi-parameter
dependence, we shall neglect the deviations from the SM for the
diagonal Higgs couplings.
 Furthermore, since the low-energy bounds only constrain significantly the
coupling $he\mu$, and the bounds on the vertex $h\tau l_i$ are not
really significant,
we have evaluated the  B.r. taking the parameter
$\eta_{\tau l_i}=0.1$;  the results are shown in
Fig.~\ref{fig1},
and one can notice that $B.r.(h\to \tau\mu)\simeq 0.01$
for $m_h\simeq 100-160$ GeV.

The mode $h\to \tau e$ has a similar branching ratio,
whereas  $B.r.(h\to e\mu)$ can reach at most values
of the order $\simeq 10^{-5}$.

\subsection{Detection of LFV Higgs decays at hadron colliders}
In order to study these LFV Higgs decays at
future hadron colliders, we shall focus on the mode $h^0 \to \tau\mu$
which, having the muon in the final state, seems easier to separate
from the backgrounds. Then, to derive bounds on the
LFV parameters,  we shall introduce the parameter $\kappa_{\mu\tau}$,
in such a way that the branching ratio for the decay
$h^0 \to \tau\mu$, is given by:
\begin{equation}
 B.r. ( h^0 \to \tau \mu)= 2 \kappa^2_{\tau\mu}
 B.r.(h_{SM} \to \tau^+\tau^-)
\end{equation}
where the dependence on the parameters $\eta_{ij}$
has been absorbed into the couplings
$\kappa_{\tau\mu}$.

 Thus, given that we naturally expect this mode to have a b.r. in
the range $10^{-2}-10^{-1}$, it seems quite interesting to try
looking for the prospectives to detect these modes at future
colliders.
 An estimate of their feasibility at Tevatron in
ref.~\cite{Diaz-Cruz:1999xe},
was followed by a more detailed background study
in ref.~\cite{Han:2000jz}, where the LHC case was included too.
 The case of a linear colliders was studied in
ref.~\cite{Sher:2000uq}.
 In this paper, we shall also include the case of
a Very Large Hadron Collider, with a c.m. energy of 40 TeV.

In order to study the possibility to detect the LFV higgs decays,
one can use the gluon-fusion mechanism to produce a single Higgs
boson; assuming that the production cross-section is of similar
strength to the SM case, about 1.2 pb for $m_h=100$ GeV, it will
allow to produce 12,000 Higgs bosons with an integrated luminosity
of 10 $fb^{-1}$. Thus, for $B.R.(h\to \tau \mu/\tau e) \simeq
10^{-1}-10^{-2}$  Tevatron can produce 1200-120 events.
Then, to determine the detectability of the signal,
 we need to study the main backgrounds to the $h \to \tau\mu$ signal, which
are dominated by Drell-Yan tau pair and WW pair production.
 In Ref.~\cite{Han:2000jz} it was proposed to reconstruct the
hadronic and electronic tau decays, assuming the following cuts:
i) For the transverse muon and jet momentum: $p^\mu_T > m_h/5$,
$p^{\pm}_T > 10$ GeV,
ii) Jet rapidity for Tevatron (LHC): $|\eta| < 2 (2.5)$
iii) The angle between the missing transverse momentum and the
muon direction: $\phi(\mu,\pm)> 160^o$.

\begin{figure}
 \includegraphics[scale = .55, angle = -90]{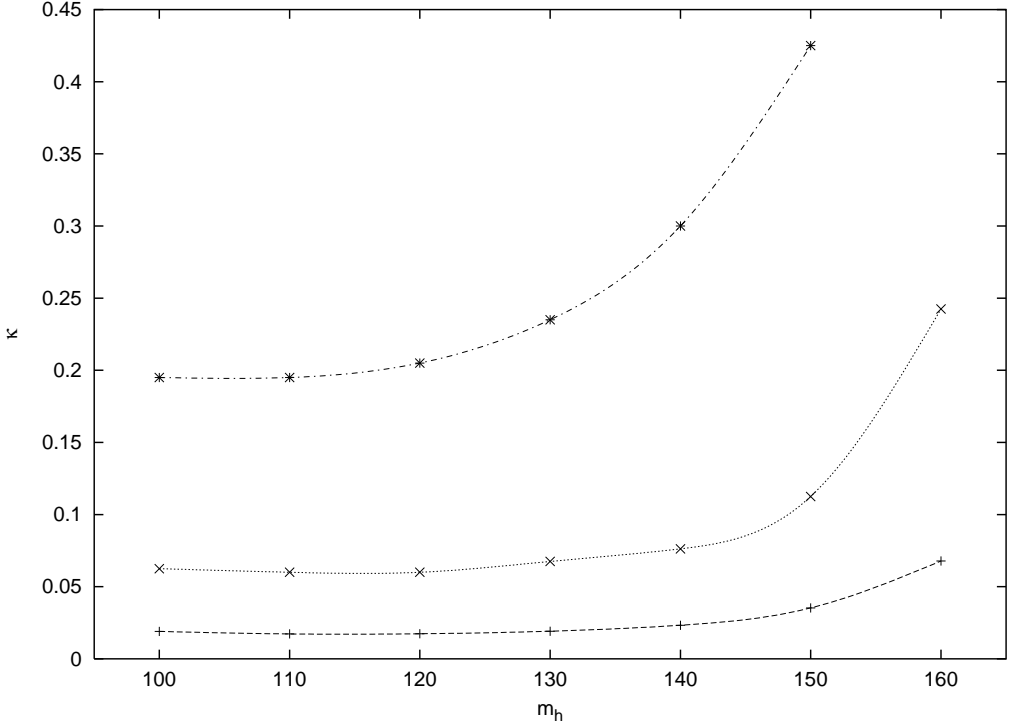}%
 \caption{Bounds on the LFV coupling $\kappa_{\tau\mu}$ that can
          be obtained at Tevatron Run-2 (dot-dashes), LHC (dots)
          and VLHC (dashes).}
 \label{fig2}
\end{figure}

The resulting bounds on the LFV higgs couplings $\kappa_{\tau\mu}$
that can be obtained at Run-2 and LHC at 95\% c.l.,
are shown in Fig.~\ref{fig2};
 one can see that it will be possible to test values
of $\kappa_{\tau\mu}$ of order 1.2-1.5 (0.15-0.2) at Tevatron
(LHC) with 4 (100) $fb^{-1}$ for $m_h = 110-130$ GeV.
In Fig.~\ref{fig2} we have also included the expected bound on
$\kappa_{\tau\mu}$ at the very large hadron collider (VLHC),
with c.m. energy of 40 TeV and integrated luminosity of
1000 fb$^{-1}$, under the very crude assumption that signal
and background can be scaled from LHC results; in this case
the sensitivity extends up to values of $\kappa_{\tau\mu}=0.1$.

\section{Conclusions}
\label{concl}

 We have studied in this paper the conditions under which flavour
violation can arise in scalar-fermion interactions,
as a result of the mixing phenomena
between the standard model and exotic fermions.
This phenomena is then discussed within the specific context
of a left-right model, which includes new fermions with mirror
properties under the new gauge group $SU(2)_R$.
Then we study phenomenological consequences of lepton
flavor violation, focusing mainly in the Higgs sector of the model.
 Bounds on the parameters of the model are obtained from
LFV processes; these results are then used to study the LFV Higgs decays
($h\to \tau l_j$ , $l_j=e,\mu$).
 We found that the LFV Higgs decays $h\to \tau \mu/\tau e$
can have large branching ratios, of order 0.01.
 The signal $h\to \tau \mu$ has good
chances to be detected at the coming stages of Tevatron
Run-II. Further studies of the signal at Tevatron Run-2 is
currently in progress ~\cite{Cotti:2002}, this time
using the realistic detector simulation of CDF.


\begin{acknowledgments}
 This work was partially supported by CONACyT, SNI and CIC-UMSNH in Mexico and
COLCIENCIAS in Colombia.
\end{acknowledgments}

\end{document}